\documentclass{article}

\usepackage[final]{neurips_2023}

\usepackage[utf8]{inputenc} % allow utf-8 input
\usepackage[T1]{fontenc}    % use 8-bit T1 fonts
\usepackage{hyperref}       % hyperlinks
\usepackage{url}            % simple URL typesetting
\usepackage{booktabs}       % professional-quality tables
\usepackage{amsfonts}       % blackboard math symbols
\usepackage{nicefrac}       % compact symbols for 1/2, etc.
\usepackage{microtype}      % microtypography
\usepackage{xcolor}         % colors
\usepackage{amsmath}
\usepackage{multirow}
\usepackage{subcaption}
\usepackage{lipsum}
\usepackage{fancyhdr}       % header
\usepackage{graphicx}       % graphics
\graphicspath{{media/}}     % organize your images and other figures under media/ folder
\usepackage{todonotes}
\usepackage{caption}
\usepackage{subcaption}
\usepackage{enumitem}

\title{
    Machine Learning for Energy-Performance-aware Scheduling
}

\author{
  Zheyuan Hu$^\dagger$, Yifei Shi$^\dagger$ \thanks{$^\dagger$ denotes equal contribution.}\\
  Department of Computer Science and Technology, \\ 
  University of Cambridge, Cambridge, UK. \\
  {\tt \small  \{zh369, ys690\}@cam.ac.uk
  }
}

\begin{document}

\maketitle

\begin{abstract} {
\itshape In the post-Dennard era, optimizing embedded systems requires navigating complex trade-offs between energy efficiency and latency. Traditional heuristic tuning is often inefficient in such high-dimensional, non-smooth landscapes. In this work, we propose a Bayesian Optimization framework using Gaussian Processes to automate the search for optimal scheduling configurations on heterogeneous multi-core architectures. We explicitly address the multi-objective nature of the problem by approximating the Pareto Frontier between energy and time. Furthermore, by incorporating Sensitivity Analysis (fANOVA) and comparing different covariance kernels (e.g., Matérn vs. RBF), we provide physical interpretability to the black-box model, revealing the dominant hardware parameters driving system performance. 
} 
\end{abstract}
    \section{Introduction} \label{sec:intro}

\par The end of Dennard scaling \cite{dennard2003design} has ushered in an era of extreme architectural heterogeneity, where modern processors integrate diverse computational units within a single silicon die. In this regime, the mapping between operating system scheduling decisions (e.g., task placement, time slice granularity, frequency scaling) and the resulting system metrics (e.g., energy consumption, instruction throughput) exhibits complex, non-linear sensitivities. Consequently, traditional heuristic schedulers often fail to navigate this landscape. While high-fidelity simulation offers a detailed evaluation platform, relying on exhaustive search or random sampling within such environments is often computationally intractable due to the combinatorial explosion of the configuration space.

\par Conversely, recent attempts to employ Deep Reinforcement Learning (DRL) \cite{khadivi2025deep} offer a data-driven alternative but trade one opacity for another. Treating the scheduler as a black-box neural network requires prohibitive training data and, crucially, offers minimal insight into \textit{why} a particular schedule is optimal. While powerful, DRL merely replaces one opacity with another, requiring prohibitive amounts of training data while offering minimal insight into \textit{why} a particular schedule is optimal.

\par In contrast to data-hungry DRL, Bayesian Optimization (BO) has demonstrated superior sample efficiency in system tuning, exemplified by works like \textit{CherryPick} \cite{alipourfard2017cherrypick} and \textit{OtterTune} \cite{van2017automatic}. However, these approaches typically target the smooth, continuous landscapes of cloud computing, overlooking the discrete discontinuities inherent to embedded heterogeneous architectures. Bridging this gap, our work focuses on the offline configuration of post-Dennard multi-core systems. We move beyond simple parameter tuning to address a fundamental scientific question: \textit{What is the latent topological structure of the optimization landscape in heterogeneous architectures, and can a data-driven framework decipher this structure to reveal interpretable, physics-compliant scheduling principles?}

\par We answer this by rethinking scheduling through the lens of \textit{Bayesian Statistics}. Rather than viewing the system merely as a black box to be optimized, we employ the Gaussian Process (GP) \cite{rasmussen2005gaussian} as a probabilistic probe to map the unknown performance function. This approach frames scheduling as a sequential model-based exploration that rigorously balances the exploration of unsampled configurations with the exploitation of efficient states. In particular, we integrate Sensitivity Analysis \cite{oakley2004probabilistic} directly into the loop, allowing us to demystify the optimization process and extract physical interpretability from the learned model.

\par The contributions of our work are summarized as follows:

\begin{enumerate}[left=0.2cm]
    \item \textbf{Methodological Validation for Non-Smooth Landscapes:} We demonstrate that the scheduling landscape is inherently non-smooth. Through rigorous kernel benchmarking, we establish that the Matérn 5/2 kernel outperforms the standard RBF kernel, as it correctly models the sharp performance cliffs associated with discrete core allocations.
    \item \textbf{Discovery of "Race-to-Idle" Physics:} By analyzing the energy-latency trade-off, our model autonomously rediscovers the "Race-to-Idle" phenomenon. We provide empirical evidence that activating high-frequency big cores often yields superior energy efficiency compared to leakage-prone low-frequency execution.
    \item \textbf{Structural Decoupling of Heterogeneity:} Our multi-objective analysis reveals a functional decoupling in hardware resources. The optimizer learns to map latency-critical tasks to big cores while leveraging little cores for energy conservation, effectively disentangling the conflicting objectives of the heterogeneous system.
\end{enumerate}

% The contributions of our work include (1) xxxx; (2) yyyy; (3) zzzz.
    % \section{Related works} \label{sec:related-work}

% % Can merge some of the system literatures.

% % Research papers that are related to our experimental designs or theory behind.

% \paragraph{Processor scheduling.}  Modern hardware or operating system designers deal with task assignment for a high degree of parallelism or low battery usage \cite{CPU_sched_Heliyon}. Classic algorithms, e.g., FIFO, Round robin, and priority-based, provide decent practical usage with strong theoretical support. However, it remains an important question to pick an appropriate time quantum, or other parameters for schedulers. They are influenced by specific workload and system settings.

% \paragraph{Heterogeneous multi-core.} Unlike homogeneous systems that deploy identical cores, heterogeneous systems are designed for multi-objective optimisation. In practice, Arm \texttt{big.LITTLE} \cite{arm_biglittle} produces two types of processors, where one for performance oriented with high frequency and the other for limited energy consumption with low frequency. 

% % \paragraph{Dynamic Voltage and Frequency Scaling (DVFS).} 

% % \paragraph{Idle management.} 

% \paragraph{Machine learning in Computer System.}  Deep Reinforcement Learning (DRL) has been adapted for scheduling \cite{khadivi2025deep}. Despite its ability to handle complex, high-dimensional problems, it lacks explainability and performs worse in unseen scenarios. We will adopt hybrid machine learning approach to incorporate domain knowledge.

\section{Related works} \label{sec:related-work}

% Can merge some of the system literatures.

% Research papers that are related to our experimental designs or theory behind.

\paragraph{Processor scheduling.} Modern hardware and operating system designers deal with task assignment to achieve a high degree of parallelism or low battery usage \cite{CPU_sched_Heliyon}. Classic algorithms, e.g., FCFS, Round Robin, and priority-based schedulers, provide decent practical usage with strong theoretical support. However, determining appropriate parameters (e.g., time quantum) remains challenging as they are heavily influenced by specific workload characteristics and system settings.

\paragraph{Heterogeneous multi-core systems.} Unlike homogeneous systems that deploy identical cores, heterogeneous systems are designed for multi-objective optimization. In practice, the Arm \texttt{big.LITTLE} architecture \cite{arm_biglittle} integrates performance-oriented high-frequency cores with energy-efficient low-frequency cores. This heterogeneity introduces a complex, non-smooth parameter space that challenges traditional heuristic tuning methods.

\paragraph{Machine learning for systems.} Recently, machine learning has been increasingly adopted to solve combinatorial system optimization problems. 
Deep Reinforcement Learning (DRL) has demonstrated remarkable capability in mapping computational graphs to heterogeneous devices, as demonstrated by Mirhoseini et al. \cite{mirhoseini2017device}.
However, despite DRL's ability to handle high-dimensional problems, it often suffers from poor sample efficiency and a lack of interpretability \cite{khadivi2025deep}. As more sample-efficient alternatives, BO and probabilistic programming have gained traction \cite{snoek2012practical}. Dalibard et al. \cite{dalibard2017boat} incorporated structural priors into BO to accelerate convergence. Similarly, Shao et al. \cite{shao2022tensor} demonstrated that leveraging probabilistic programs for tensor program optimization can significantly outperform traditional search strategies in compiler tuning. Our work aligns with this trend of sample-efficient optimization but specifically focuses on the \textit{kernel selection} mechanism within BO. 
    \section{Methodology}  \label{sec:method}
In this section, we define how our simulator is designed, including tasks, processors and schedulers, followed by the turnaround time, priority-aware latency, energy metrics with respect to processor frequency. Motivated by real-world problems, we aim to grasp knowledge with the help from machine learning techniques.

\subsection{Simulator design} \label{sec:simulator-design}

\begin{figure}[htbp]
  \centering
  \begin{subfigure}[b]{0.44\linewidth}
    \centering
    \includegraphics[width=\linewidth]{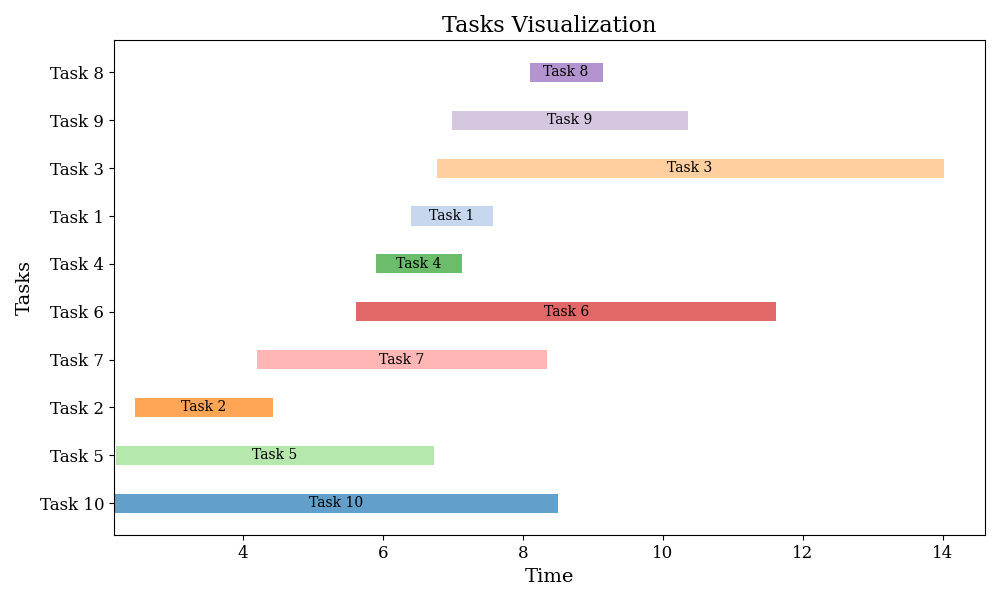}
    \caption{Tasks factory visualisation.}
    \label{fig:task_factory_visualization}
  \end{subfigure}
  \hfill
  \begin{subfigure}[b]{0.54\linewidth}
    \centering
    \includegraphics[width=\linewidth]{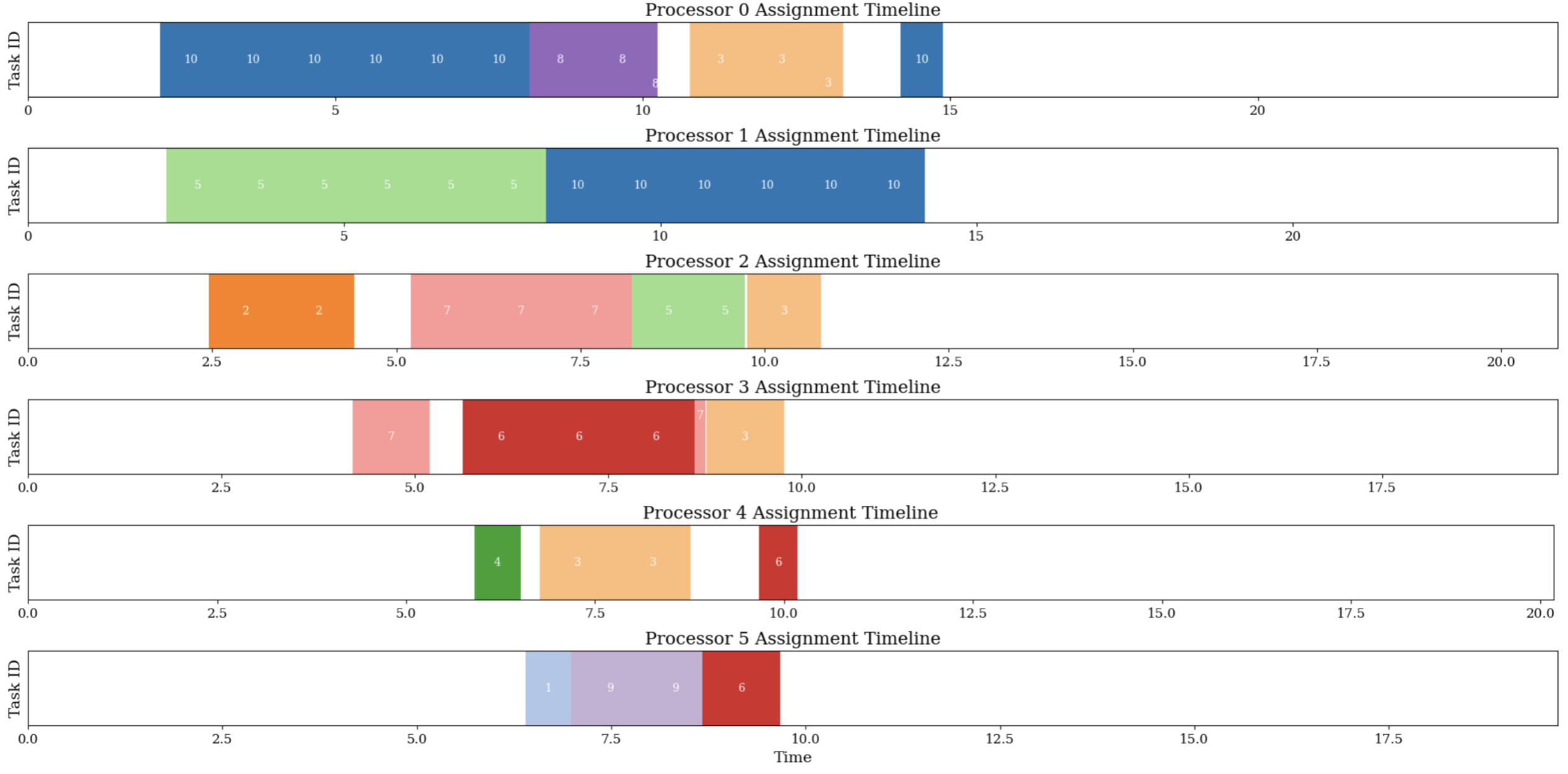}
    \caption{Round robin for six heterogeneous processors.}
    \label{fig:RR_sched_6_cores}
  \end{subfigure}
  \caption{Simulator visualisation and results.}
  \label{fig:simulator_visualisation_results}
\end{figure}

\paragraph{Task definition.} The system consists of a set of tasks $\mathcal{T} = \{\tau_i\}_{i=1}^{N_{\text{task}}}$. Each task $\tau$, e.g., a thread or process, can be defined as a five-element tuple: $\tau = (t_a, t_f, p, N_{\text{IC}}, E), \quad t_a, t_f, E \in \mathbb{R},\; p, N_{\text{IC}} \in \mathbb{Z}$,
where arrival time $t_a$, preferred finish time $t_f$, instruction count $N_{\text{IC}}$, priority level $p$, and energy consumed $E$ are task attributes. These are summarised in Table \ref{tbl:task_attribute}. For each task in the given set, its attributes are randomly initialised over a fixed region by a data factory class (Figure \ref{fig:task_factory_visualization}). The factory method pattern creates task objects following the Open-Closed Principle (OCP). As an example, the tasks follow the Poisson process and the waiting time modelled by exponential distribution. Regarding detailed performance analysis, please refer to § \ref{sec:sim-metric}.

\begin{table}[htbp]
\centering
\caption{Task attributes and their data type.}
\label{tbl:task_attribute}
\begin{tabular}{lcccccc}
\hline
 & task id & arrival time & instruction count & priority & energy & finish time \\
\hline
Type & int & float & float & int & float & Optional[float] \\
\hline
\end{tabular}
\end{table}

\paragraph{Heterogeneous multi-core design.} The system consists of a set of processors $\mathcal{P} = \{P_i\}_{i=1}^{N_{\text{cores}}}$. Each processor is set to a fixed frequency during the whole execution time for simplicity, i.e. Dynamic Voltage and Frequency Scaling, or DVFS for short, is not considered. 

In this work, we aim to investigate how many types are needed for a given workload and how frequencies should be set for each processor type. There is a balance between high performance and low energy consumption. This is important for configuring per-core CPU frequency settings.

\paragraph{Scheduler design.} The core goal of schedulers is to dispatch tasks to appropriate processors. There are various task assignment algorithms in the literature, which differ by their simplicity, overhead, fairness, and whether they are preemptive. We implement the following scheduling methods (also shown in Table \ref{tbl:scheduling-comparison}),

\begin{itemize}[left=0.2cm]
  \item First-Come, First-Served (FCFS), also known as First-In, First-Out (FIFO). Tasks are executed strictly according to the order of arrival time. The pros are its predictability, starvation-free. However, short tasks that arrive late will wait longer time, which is termed the Convoy effect.
  \item Round Robin (RR) is a specific implementation of  time-sliced or quantum based scheduling (Figure \ref{fig:RR_sched_6_cores}), where tasks are preempted from the processors after the quantum used up. This saves computation time for short tasks, but keeps the average turnaround time larger.
  \item Priority-based scheduling. Tasks are dispatched based on the priority level. It can either be preemptive or non-preemptive. It favors high-priority tasks, but might starve those low-priority processes.
\end{itemize}

\begin{table}[htbp]
\centering
\caption{Comparison of FCFS, Round Robin, and priority-based scheduling algorithms.}
\label{tbl:scheduling-comparison}
\begin{tabular}{|l|c|c|c|}
\hline
\textbf{Features} & \textbf{FCFS} & \textbf{Round Robin (RR)} & \textbf{Priority-based} \\
\hline
decision basis & arrival time & time quantum & priority level \\
preemptive       & $\times$ & $\checkmark$ & $\circ$ \\
starvation       & $\times$ & $\times$  & $\checkmark$  \\
context switchg & $\downarrow$ & $\uparrow$ & $\circ$  \\
use case         & batch processing & time-sharing & real-time / critical \\
% Fairness         & Low & High & Medium \\
% Response Time    & Poor & Good & Good (high priority) \\
% Implementation   & Very Simple & Moderate & Moderate \\
\hline
\end{tabular}
\end{table}

In this project, we aim to learn which scheduler works best for a given series of tasks, including their attributes, e.g., time quantum for Round Robin, in the context of energy-performance-aware scheduling.

\paragraph{Simulation environment and logging.} We adopt \texttt{SimPy} \cite{simpy}, a process-based discrete-event framework, as our backend simulation environment. It supports waiting, interrupting a process and shared resources. Our simulator is divided into \texttt{Task}, \texttt{Processors},  \texttt{Schedulers}, and \texttt{Simulators} packages, where details are described as above.  For efficient verification and debugging, we deploy \texttt{logging} to track every event of the simulation.

An example log output generated via \texttt{logging}.
\begin{verbatim}
INFO:Task factory:Created <N> tasks. Processor <p_id>:Frequency = <f>.
INFO:Round Robin scheduler:Initialized with quantum = <q>.
INFO:Simulator:Task <t_id> arrived at time <t>, with instruction <count>.
INFO:Round Robin scheduler:Task <t_id> quantum expired, enqueuing again.
\end{verbatim}

% INFO:simulator:Simulator environment created.
% INFO:Priority scheduler:Task <id> added to scheduler queue at time <t>.
% INFO:Priority scheduler:Scheduling Task <id> on Resource <r> at time <t>.
% INFO:processor <id>:started processing Task <id> with priority <p> at time <t> with remaining size <s>.
% INFO:processor <id>:finished processing Task <id> at time <t>.

\subsection{Main objectives} \label{sec:sim-metric}
\paragraph{Latency and frequency modelling.} Instructions per Cycle (IPC) is governed by architectural design, e.g., superscalar pipelines, out-of-order execution, speculation, and branch prediction. We assume that IPC is a fixed parameter, not subject to change or variation in our analysis. For a task with instruction count $N_{\text{IC}}$, the number of cycles it takes $N_{\text{cycle}} = \frac{N_{\text{IC}}}{\text{IPC}}$ and the total execution time, also known as turnaround time is $\quad t = \frac{N_{\text{cycle}}}{f} = \frac{N_{\text{IC}}}{\text{IPC} \cdot f} \propto \frac{1}{f}$, where $f$ is the operating frequency in cycles per second, or Hertz (Hz). 

\paragraph{Priority-aware latency.} Each task is categorized into levels of importance $p$, as in the standard Operating System. When aggregating the latency, those with higher priority are given a larger weight: $
w(p) = \frac{1}{(p + 1)^2}, \quad p \in \{0, 1, 2, \dots, K\}$, where a lower $p$ stands for higher priority and $K$ is the least priority level.

\begin{figure}[htbp]
  \centering
  \begin{subfigure}[b]{0.45\linewidth}
    \centering
    \includegraphics[width=\linewidth]{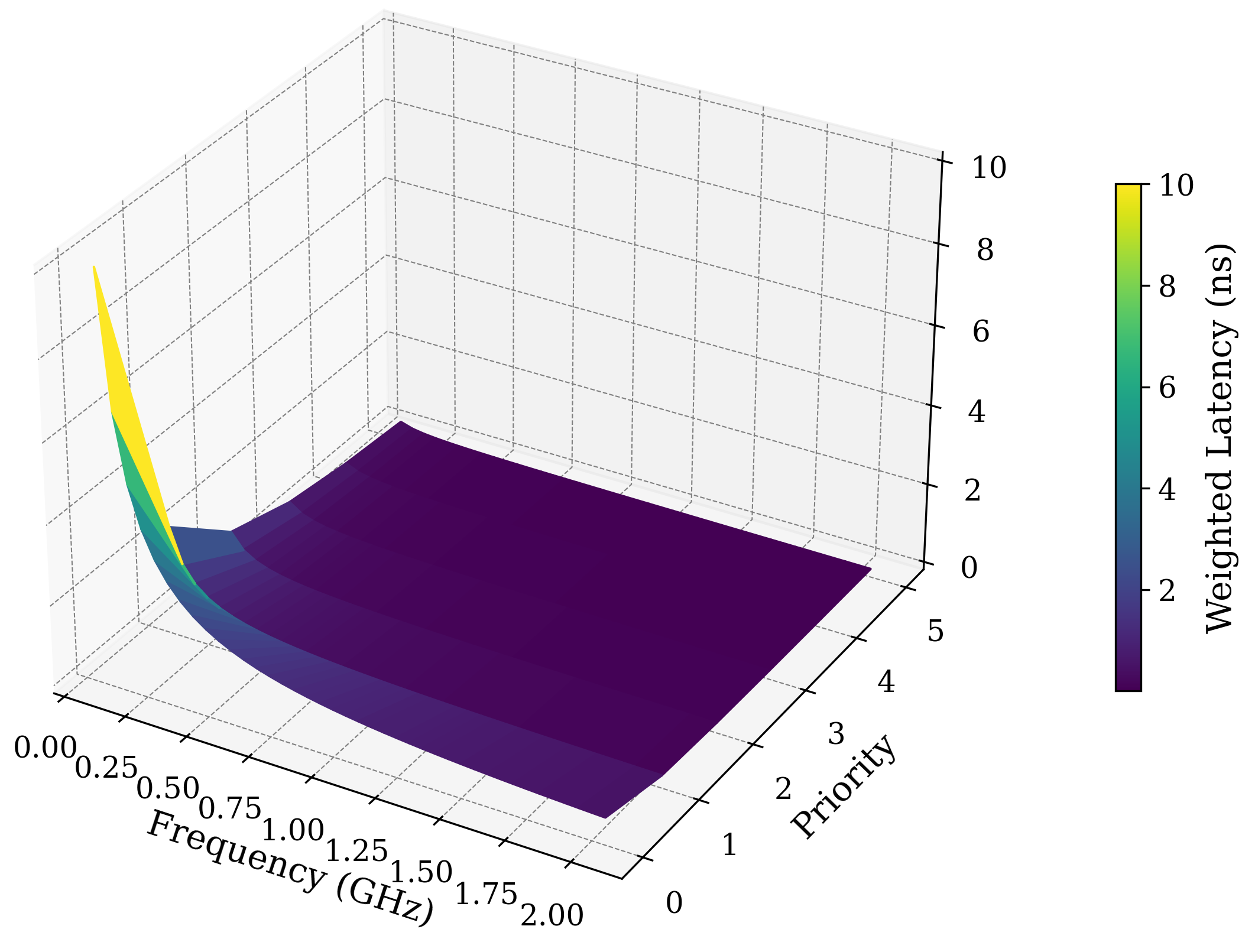}
    \caption{Weight latency vs. frequency and task priority.}
    \label{fig:weighted_latency_by_freq_priority}
  \end{subfigure}
  \hfill
  \begin{subfigure}[b]{0.45\linewidth}
    \centering
    \includegraphics[width=\linewidth]{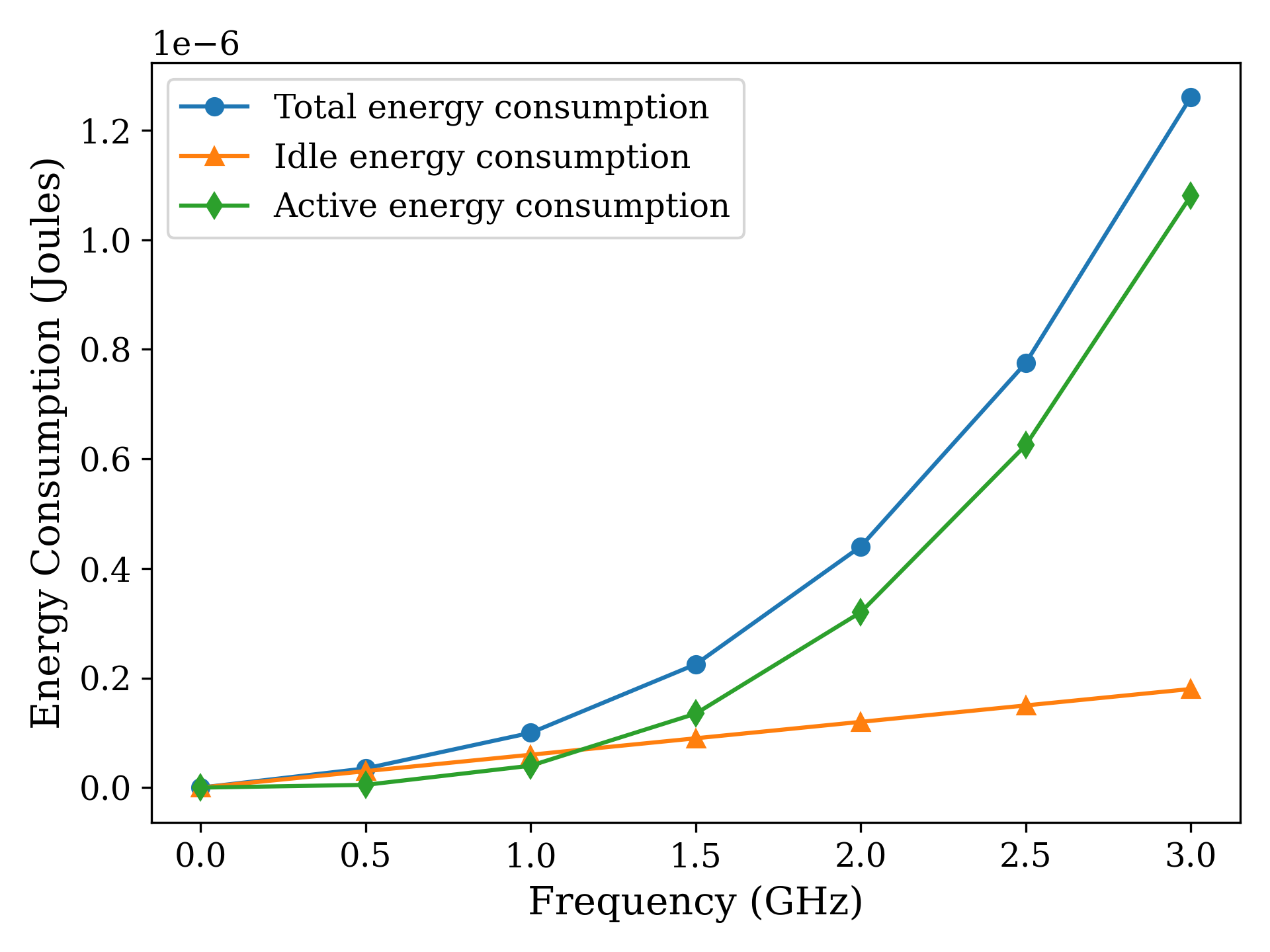}
    \caption{Energy consumption vs. frequency.}
    \label{fig:energy_consumption_vs_freq}
  \end{subfigure}
  \caption{Processor energy consumption and task latency.}
  \label{fig:energy_latency_plot_vs_freq_priority}
\end{figure}

The priority-aware aggregated latency for a dataset of tasks is the weighted average given by $t_{\text{aggregated}} = \sum_{i=1}^{N_{\text{task}}} w_i(p)  \cdot t_i$, where $N_{\text{task}}$ is the total number of tasks, $t_i$ is per-task latency, and $w_i(p)$ is the weight for each task. Figure \ref{fig:weighted_latency_by_freq_priority} shows the weighted task latency as a function of both processor frequency and task priority.

\paragraph{Energy and frequency modelling.} The total power of the computer systems consists of both the processor and the rest (e.g., display, memory subsystems, sensors). Here we focus on the power consumption $P_{\text{XPU}}$ of the processors XPU, encompassing CPU, GPU, NPU and other specialised ASIC. Here we focus on two parts \cite{de2013energy}, i.e. dynamic and leakage power: $P_{\text{XPU}} = P_{\text{dynamic}} + P_{\text{leakage}}$.

The \textit{dynamic power} $P_{\text{dynamic}}$ is that consumed when transistors switch states, i.e. from 0 to 1 or vice versa. This comprises the key and basic step of processors for nearly all computation. With $\alpha \in [0,1]$ of the processors actively switching, the dynamic power is, $P_{\text{dynamic}} = \alpha \; C \cdot V^2 \cdot f$, where $C$ is the system capacitance as measured in the unit of Farad (F), $V$ is the applied voltage expressed in Volts (V), and $f$ is the operating frequency in cycles per second, or Hertz (Hz). When there is no computation taking place, which means the processor is idle, the dynamic power is zero, i.e. $P_{\text{dynamic}}=0$.

The \textit{leakage power} $P_{\text{leakage}}= V \cdot I_{\text{leakage}}$ captures power from flowed current $I_{\text{leakage}}$ when Metal Oxide Semiconductor Field Effect Transistor, or MOSFET for short, is in either ON or OFF state, which occurs when the gate-source voltage $V_{GS}$ is above or below the threshold voltage of the MOSFET $V_{TH}$, i.e. $V_{GS} \geq V_{TH}$ or $V_{GS} < V_{TH}$.

Note that leakage power occurs even when the processor is not actively processing tasks and is not turned OFF, which prevents all transistors from turning ON at all times. This leads to a phenomenon termed the power wall in the semiconductor world.

The achievable frequency is constrained by gate delay. Here we assume that the operating voltage is far from the threshold $V_{TH}$ and thus can be approximated by a linear dependency with respect to the operating frequency,
\begin{equation}
f = \frac{1}{t_{\text{delay}}} \propto \frac{(V-V_{th})^\gamma}{V} \approx k_V V + b_f, \quad \text{ where }\gamma \in \mathbb{R} \subset [1, 2], \;  k_V, b_f \in \mathbb{R}.
\end{equation}
The relationship between \textit{energy} $E$ (Joules) and \textit{power} $P$ (Joules per second or Watts) over the time period $\tau$ is $E(\tau) = \int_0^\tau P(t) \; dt = P \cdot \tau$, if the power is constant along execution time.

Thus, the total processor energy $E$ over a fixed time period $\tau$ with respect to frequency $f$ is, when $b_f=0$, as shown in Figure \ref{fig:energy_consumption_vs_freq}, including the active, idle and total energy as a function of the processor frequency over a fixed period of time,
\begin{equation}
E(f) = P \cdot \tau
= (\alpha C\cdot \frac{1}{k^2_V} f^3 + I_{\text{leakage}}  \frac{1}{k_V} f ) \cdot \tau.
\end{equation}

Likewise, for tasks with a fixed number of instructions $N_{\text{IC}}$, the total processor energy $E$ is,
\begin{equation}
E(f) = P \cdot \frac{N_{\text{IC}}}{f \cdot \text{IPC}} 
= (\alpha C\cdot \frac{1}{k^2_V} f^3 + I_{\text{leakage}}  \frac{1}{k_V} f ) \cdot \frac{1}{f}
= \alpha C\cdot \frac{1}{k^2_V} f^2 + I_{\text{leakage}}  \frac{1}{k_V}.
\end{equation}

% \paragraph{Experiments setup.} Moved to sec:Experiments setup

\subsection{Emulator Design}

% interesting kernels, or add, multiplication of some kernels.
\paragraph{Gaussian Process Priors}
To model the complex, non-linear landscape of CPU scheduling performance, we employ Gaussian Process regression as a probabilistic surrogate \cite{rasmussen2005gaussian}. A GP defines a distribution over functions, specified by a mean function $m(\mathbf{x})$ (assumed zero after normalization) and a covariance kernel $k(\mathbf{x}, \mathbf{x}')$. The kernel encodes our prior assumptions about the system's smoothness. 

\par We explicitly compare the \textit{Radial Basis Function (RBF)} kernel and the \textit{Matérn} kernel. The RBF kernel assumes infinite differentiability, which implies a highly smooth objective landscape \cite{genton2002classes}. However, CPU scheduling often involves discrete parameters (e.g., core counts) and performance cliffs (e.g., sudden latency spikes when buffers overflow). 

\par Therefore, we adopted the \textit{Matérn 5/2 kernel} ($\nu=2.5$) as the optimal compromise. This kernel relaxes the infinite smoothness assumption of RBF to capture system non-linearities (e.g., performance cliffs) while retaining sufficient differentiability (twice differentiable) to ensure robust numerical stability and efficient gradient-based optimization of the acquisition function.

\paragraph{Bayesian Optimization and Acquisition Strategy}
Based on the GP posterior, BO iteratively selects the next configuration $\mathbf{x}_{t+1}$ by maximizing an acquisition function $a(\mathbf{x})$ that balances exploration (variance reduction) and exploitation (mean minimization) \cite{snoek2012practical, shahriari2015taking}.
\begin{itemize}[left=0.2cm]
    \item Single-Objective (LogEI) \cite{ament2023unexpected}: For scalarized cost minimization, we employ \textit{Log Expected Improvement} (LogEI). While conceptually equivalent to standard EI, LogEI performs computations in the logarithmic space to prevent numerical underflow in regions of high posterior probability. This ensures robust gradient estimation even when the expected improvement is asymptotically small near the global optimum \cite{wilson2018maximizing}.
    \item Multi-Objective (Pareto Frontier) \cite{deb2011multi}: For the Pareto optimization phase, we utilize the \textit{Expected Hypervolume Improvement} (EHVI)  \cite{daulton2020differentiable}. Unlike scalarization methods, EHVI explicitly targets the increase in the hypervolume dominated by the current Pareto front $\mathcal{P}^*$. By integrating over the non-dominated region of the posterior distribution, this strategy effectively incentivizes the discovery of diverse trade-off solutions that extend the frontier towards lower energy and latency.
\end{itemize}

\paragraph{Hyperparameter Sensitivity Analysis}
To interpret the black-box simulation model and identify the dominant factors driving system performance, we conduct a post-hoc sensitivity analysis on the fitted surrogate models \cite{oakley2004probabilistic}. Since our surrogate model is anisotropic, the model learns a distinct \textit{length-scale parameter} $l_d$ for each input dimension $d$. 
There is an inverse relationship between length-scale and sensitivity: $\text{Importance}_d \propto \frac{1}{l_d}$.

\par A small length-scale $l_d$ implies that the objective function changes rapidly along dimension $d$, indicating high sensitivity. Conversely, a large length-scale suggests the dimension is irrelevant (the function is flat along that axis).

\par In the multi-objective context, we perform this analysis separately for the Energy and Time objectives. This allows us to decouple conflicting drivers, revealing, for example, which hardware resources constitute the bottleneck for latency versus which are the primary sources of power consumption.

\subsection{Evaluation Metrics}
\par To comprehensively assess the performance of the proposed BO framework, we utilize a combination of scalarized cost metrics, multi-objective indicators, and statistical distribution analysis.

\paragraph{Weighted Log-Sum Cost}
For the single-objective phase, we employ a scalarization technique to transform the vector-valued outputs into a single minimization target. We apply a logarithmic transformation to handle the disparate magnitudes of energy ($E$) and time ($T$):
\begin{equation}
\mathcal{L}(\mathbf{x}) = \beta \cdot \ln(E(\mathbf{x})) + \gamma \cdot \ln(T(\mathbf{x}))
\label{eq:metric}
\end{equation}
where $\beta$ and $\gamma$ are user-defined preference weights. This metric evaluates the optimizer's ability to find a specific trade-off point on the Pareto frontier.

\paragraph{Pareto Optimality and Hypervolume Indicator}
In the multi-objective phase, we aim to approximate the true \textit{Pareto Front} $\mathcal{P}^*$. A solution $\mathbf{x}_1$ dominates $\mathbf{x}_2$ ($\mathbf{x}_1 \prec \mathbf{x}_2$) if it is strictly better in at least one objective and no worse in others.
To quantitatively evaluate the quality of the obtained Pareto set, we use the \textit{Hypervolume (HV) Indicator} \cite{daulton2020differentiable}. HV measures the volume of the objective space dominated by the solution set $\mathcal{P}$ relative to a reference point $\mathbf{r}$. A larger Hypervolume indicates a better approximation of the true Pareto front, reflecting both the \textit{convergence} (closeness to the optimum) and \textit{diversity} (spread) of the solutions.

% % TODO: Maybe delete this part if no EDF involved in results section
% \paragraph{Empirical Distribution Function (EDF)}
% To assess the robustness and consistency of the optimizer across multiple trials, we analyze the \textit{Empirical Distribution Function (EDF)} of the objective values. The EDF, denoted as $F_N(v)$, represents the fraction of trials where the achieved objective value is less than or equal to a threshold $v$:
% \begin{equation}
% F_N(v) = \frac{1}{N} \sum_{i=1}^{N} \mathbb{I}(y_i \leq v)
% \end{equation}
% where $\mathbb{I}(\cdot)$ is the indicator function and $N$ is the total number of evaluations. A steeper EDF curve (rising quickly near the minimum value) indicates a higher probability of finding high-quality solutions efficiently.

\paragraph{Baselines}
We compare the performance of our Gaussian Process-based BO against a \textit{Random Search} baseline. Random Search serves as a fundamental benchmark to demonstrate the sample efficiency and convergence speed of the active learning strategy employed by our BO framework.
    \section{Experiments}
\label{sec:exp}

\subsection{Experiment Setup}

% We evaluate our proposed framework using a custom event-driven CPU scheduling simulator implemented in Python. 
The processor constants and the units of time and frequency are set according to Table \ref{tbl:processor-const}, in accordance with the range specified in the standard processor design document. All experiments were conducted on a workstation with 9 CPU Cores and 18 GiB RAM, using \texttt{Optuna} \cite{akiba2019optuna} and \texttt{BoTorch} \cite{balandat2020botorch} libraries.

\begin{table}[htbp]
\centering
\caption{Constants used in the processor power and frequency model.}
\label{tbl:processor-const}
\begin{tabular}{cccccc}
\hline
$k_V$ & $b_f$ & $C$ & $I_{\text{leakage}}$ & unit of $t$  & unit of $f$ \\
\hline
$5 \times 10^{9}\,\mathrm{Hz/V}$ &
$0\,\mathrm{Hz}$ &
$1 \times 10^{-9}\,\mathrm{F}$ &
$3 \times 10^{-1}\,\mathrm{A}$ &
ns ($\times 10^{-9}$ s) &
GHz ($\times 10^{9}$ Hz) \\
\hline
\end{tabular}
\end{table}

\paragraph{Simulation Environment}
The workload is generated using a Poisson process with an average arrival rate $\lambda$. Unless otherwise stated (e.g., in the robustness analysis), the default simulation parameters are fixed as follows: (1) Total Simulation Time: $T_{end} = 1000$ ms, (2) Total Tasks: $N_{tasks} = 500$, (3) Default Workload Pressure: $\lambda = 1.0$ (tasks/ms).
% \begin{itemize}[left=0.2cm]
%     \item Total Simulation Time: $T_{end} = 1000$ ms.
%     \item Total Tasks: $N_{tasks} = 500$.
%     \item Default Workload Pressure: $\lambda = 1.0$ (tasks/ms).
% \end{itemize}

\paragraph{Design Space (Search Space)}
The optimization problem involves a mixed-variable search space $\mathcal{X}$, consisting of continuous (frequency, time quantum), integer (core counts), and categorical (scheduling policy) parameters. The bounds for each decision variable are detailed in Table~\ref{tab:search_space}. The \textit{Time Quantum} parameter is conditional and only active when the scheduling strategy is non-preemptive (i.e., Round Robin or Priority).

\begin{table}[h]
    \centering
    \caption{Definition of the Search Space $\mathcal{X}$. }
    \label{tab:search_space}
    \begin{tabular}{@{}llcc@{}}
    \toprule
    \textbf{Component} & \textbf{Parameter} & \textbf{Type} & \textbf{Range / Options} \\ \midrule
    \multirow{2}{*}{Little Cores} & Frequency (GHz) & Continuous & $[0.5, 1.5]$ \\
     & Count & Integer & $\{0, 1, \dots, 4\}$ \\ \midrule
    \multirow{2}{*}{Medium Cores} & Frequency (GHz) & Continuous & $[1.0, 2.5]$ \\
     & Count & Integer & $\{0, 1, \dots, 4\}$ \\ \midrule
    \multirow{2}{*}{Big Cores} & Frequency (GHz) & Continuous & $[1.5, 3.5]$ \\
     & Count & Integer & $\{0, 1, \dots, 4\}$ \\ \midrule
    \multirow{2}{*}{Scheduler} & Strategy & Categorical & $\{\text{FCFS}, \text{RR}, \text{Priority}\}$ \\
     & Time Quantum (ms) & Continuous & $[0.5, 5.0]$ \\ \bottomrule
    \end{tabular}
\end{table}

\paragraph{Optimization Configuration}
To ensure statistical robustness, each optimization experiment consists of a total budget of $N=100$ trials. 
% \par 
To initialize the Gaussian Process surrogate model, we employ a \textit{warm-up phase} of $N_{init}=10$ trials sampled using a Sobol sequence, which provides better space-filling properties than uniform random sampling.

\subsection{Experimental Design}

\par To comprehensively evaluate the proposed BO framework, we designed four distinct experimental scenarios. Each scenario targets a specific aspect of the system's behaviour and the optimizer's capability.

% 1. Kernel Selection 
\paragraph{Surrogate Model Calibration }
In this phase, we benchmark different kernels (RBF vs. Matérn) under a standard workload ($\lambda=1.0$). 
% \par 
The objective is to determine which kernel best captures the discrete and non-smooth landscape of the CPU scheduling problem. The selected kernel is then used as the baseline for subsequent experiments.

% 2. Preference Analysis
\paragraph{Preference Sensitivity Analysis}
Since the cost function $\mathcal{L}$ is a weighted sum of energy and time, system behaviour is highly sensitive to the weights $\beta$ (energy) and $\gamma$ (time). We vary these coefficients to simulate different user priorities:
(1) \textit{Performance-First:} High penalty on time ($\gamma > \beta$), (2) \textit{Energy-First:} High penalty on energy ($\beta > \gamma$),  (3) Equal weights ($\beta = \gamma$).
% \begin{itemize}[left=0.2cm]
%     \item \textit{Performance-First:} High penalty on time ($\gamma > \beta$).
%     \item \textit{Energy-First:} High penalty on energy ($\beta > \gamma$).
%     \item \textit{Balanced:} Equal weights ($\beta = \gamma$).
% \end{itemize}
% 
% \par 
The objective is to verify if the optimizer can correctly shift the hardware configuration (e.g., scaling frequencies or core counts) to align with the specified high-level preferences.

\paragraph{Workload Robustness Testing}
We evaluate the optimizer's robustness by varying the task arrival rate $\lambda$ (from low load $\lambda=0.5$ to high load $\lambda=5.0$).
% \par 
The objective is to investigate how the optimal architectural configuration evolves under pressure. Specifically, we aim to observe if the system automatically scales up resources (e.g., activating big cores) to prevent latency cliffs during peak loads.

\paragraph{Multi-Objective Pareto Exploration}
Finally, to overcome the limitations of fixed weights, we decouple the objectives and perform Multi-Objective Optimization (MOO). Instead of minimizing a scalar loss, we aim to approximate the Pareto Frontier.
% \par 
The objective is to uncover the intrinsic trade-off curve between Energy and Time, providing a set of non-dominated solutions that allow system administrators to make a posteriori decisions without manually tuning weights.
    \section{Results and Analysis}
\label{sec:res}

\par To comprehensively assess the performance of the proposed BO framework, we utilize a combination of scalarized cost metrics, multi-objective indicators, and statistical distribution analysis. Table~\ref{tab:all_results} presents a summary of the optimal configurations found across all experimental scenarios.

\begin{table}[htbp]
    \centering
    \caption{Comprehensive Summary of Optimal Configurations found by Bayesian Optimization vs. Baseline. \textbf{Panel A} compares different search strategies under the default balanced workload. \textbf{Panel B} demonstrates adaptation to user preferences. \textbf{Panel C} reveals the system's robustness strategy under varying workload intensities. Note: Objective values are only comparable within the same Panel.}
    \label{tab:all_results}
    \resizebox{\textwidth}{!}{% Auto-resize to fit page width
    \begin{tabular}{@{}llccclc@{}}
    \toprule
    \multirow{2}{*}{\textbf{Experiment Group}} & \multirow{2}{*}{\textbf{Scenario / Variant}} & \multicolumn{3}{c}{\textbf{Core Config (Count $\times$ Freq)}} & \multirow{2}{*}{\textbf{Scheduler}} & \multirow{2}{*}{\textbf{Obj. Value}} \\ \cmidrule(lr){3-5}
     &  & \textbf{Little} & \textbf{Medium} & \textbf{Big} &  &  \\ \midrule
    
    % --- Panel A: Strategy Comparison (Fixed: Added Matern 3/2) ---
    \multicolumn{7}{l}{\textit{\textbf{Panel A: Search Strategy Comparison (Balanced, $\lambda=1.0$)}}} \\ 
    \multirow{4}{*}{Algorithm} 
     & BO (Matérn 5/2) & $3 \times 1.5$ & - & $2 \times 1.5$ & FCFS & -19.65 \\
     & BO (Matérn 3/2) & - & $1 \times 1.6$ & $\mathbf{4 \times 1.5}$ & FCFS & -19.65 \\ 
     & BO (RBF) & $1 \times 1.5$ & $\mathbf{4 \times 1.3}$ & - & RR (4.1ms) & -19.63 \\ 
     & Random Baseline & $\mathbf{4 \times 1.2}$ & $2 \times 1.8$ & - & RR (2.0ms) & -19.57 \\ \midrule
     
    % --- Panel B: Metric Evaluation (Unchanged) ---
    \multicolumn{7}{l}{\textit{\textbf{Panel B: Preference Adaptation (Metric Shifts)}}} \\ 
    \multirow{3}{*}{Metric Weights} 
     & Balanced & $3 \times 1.5$ & - & $2 \times 1.5$ & FCFS & -19.65 \\
     & Energy-First & - & $2 \times 2.5$ & $\mathbf{2 \times 3.5}$ & FCFS & -18.35 \\
     & Time-First & $3 \times 0.7$ & - & $\mathbf{3 \times 1.5}$ & Priority (0.7ms) & -61.66 \\ \midrule

    % --- Panel C: Robustness Test (Unchanged) ---
    \multicolumn{7}{l}{\textit{\textbf{Panel C: Workload Robustness ($\lambda$ Variation)}}} \\ 
    \multirow{3}{*}{Arrival Rate ($\lambda$)} 
     & Low ($\lambda=0.5$) & $3 \times 1.4$ & - & - & Priority (3.8ms) & -19.59 \\
     & High ($\lambda=2.5$) & $\mathbf{4 \times 1.5}$ & $\mathbf{4 \times 1.5}$ & $\mathbf{4 \times 1.5}$ & FCFS & -19.74 \\
     & Extreme ($\lambda=5.0$) & $1 \times 0.5$ & - & - & RR (0.7ms) & -20.63 \\ \bottomrule

    \end{tabular}%
    }
\end{table}

\subsection{Kernel Selection}

\par In this section, we evaluate the impact of the Gaussian Process covariance kernel on the optimization performance. We compared the Matérn 5/2, Matérn 3/2, and RBF kernels under the standard balanced metric defined by Equation~\ref{eq:metric}, with the hyperparameter of $\beta=1, \gamma=1$.

\par
\begin{figure}[htbp]
    \centering
    \begin{subfigure}[b]{0.32\textwidth}
        \centering
        \includegraphics[width=\textwidth]{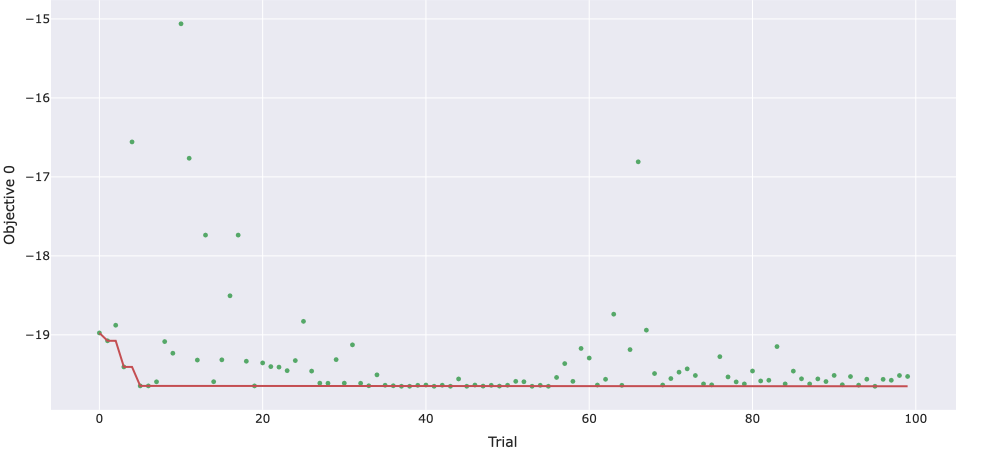}
        \caption{Optimization History for Matern 2.5.  }
        \label{fig:res_kernel_matern2.5}
    \end{subfigure}
    \hfill
    \begin{subfigure}[b]{0.32\textwidth}
        \centering
        \includegraphics[width=\textwidth]{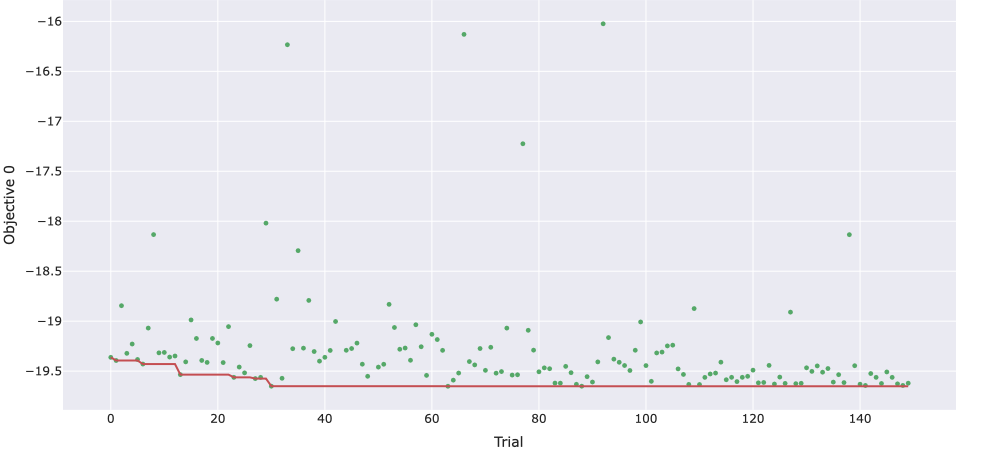}
        \caption{Optimization History for Matern 1.5. }
        \label{fig:res_kernel_matern1.5}
    \end{subfigure}
    \hfill 
    \begin{subfigure}[b]{0.32\textwidth}
        \centering
        \includegraphics[width=\textwidth]{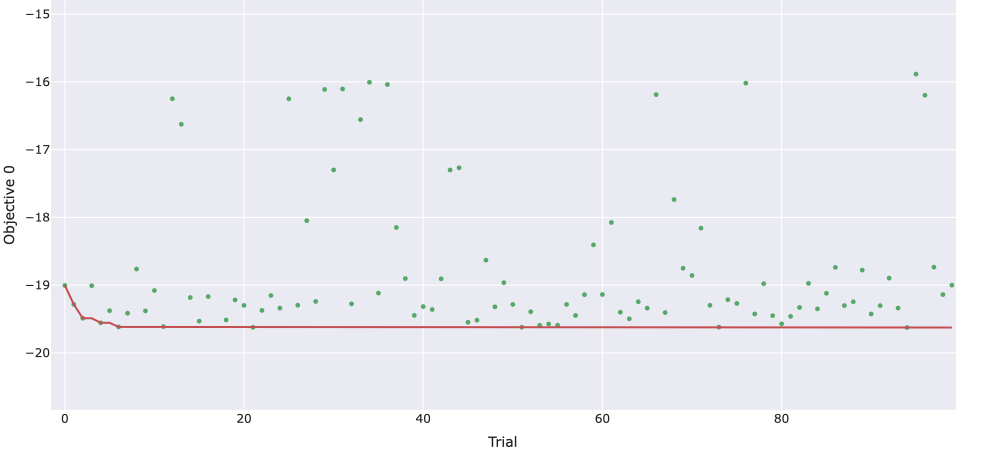}
        \caption{Optimization History for RBF. }
        \label{fig:res_kernel_rbf}
    \end{subfigure}
    
    \caption{Results of Bayesian Optimization history for different types of kernels. }
    \label{fig:res_kernel}
\end{figure}

 Figure~\ref{fig:res_kernel} illustrates the optimization history (objective value vs. trial number) for the three kernels. 
\begin{itemize}[left=0.2cm] 
\item Convergence Speed: As observed in Figure~\ref{fig:res_kernel_matern2.5}, the Matérn 5/2 kernel demonstrates the most stable convergence trajectory. It rapidly identifies the high-performance region within the first 20 trials and maintains a low variance in subsequent iterations. 
\item Impact of Smoothness: The RBF kernel (Figure~\ref{fig:res_kernel_rbf}) assumes infinite differentiability and performs slightly worse than the Matérn family. This result validates our hypothesis that the CPU scheduling landscape is non-smooth because it is characterized by discrete core counts and integer parameters. Consequently, the RBF kernel tends to over-smooth the objective function and potentially misses the sharp performance cliffs caused by discrete resource changes.
\item Roughness Handling: The Matérn 3/2 kernel (Figure~\ref{fig:res_kernel_matern1.5}) performs comparably to Matérn 5/2 but exhibits slightly higher exploration variance. 
\end{itemize}

\par Based on these results, we selected Matérn 5/2 as the baseline kernel for subsequent experiments, as it offers the optimal balance between modelling the rough landscape and maintaining numerical stability for the acquisition function.

\par
\begin{figure}[htbp]
    \centering
    \begin{subfigure}[b]{0.32\textwidth}
        \centering
        \includegraphics[width=\textwidth]{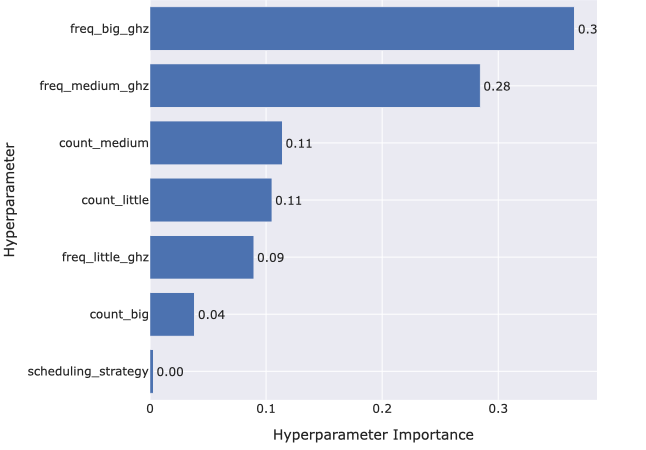}
        \caption{Sensitivity Analysis for Matern 2.5.  }
        \label{fig:analy_kernel_matern2.5}
    \end{subfigure}
    \hfill
    \begin{subfigure}[b]{0.32\textwidth}
        \centering
        \includegraphics[width=\textwidth]{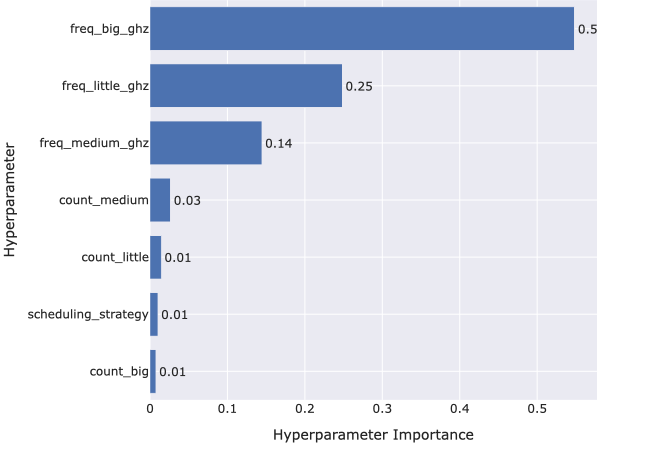}
        \caption{Sensitivity Analysis for Matern 1.5. }
        \label{fig:analy_kernel_matern1.5}
    \end{subfigure}
    \hfill 
    \begin{subfigure}[b]{0.32\textwidth}
        \centering
        \includegraphics[width=\textwidth]{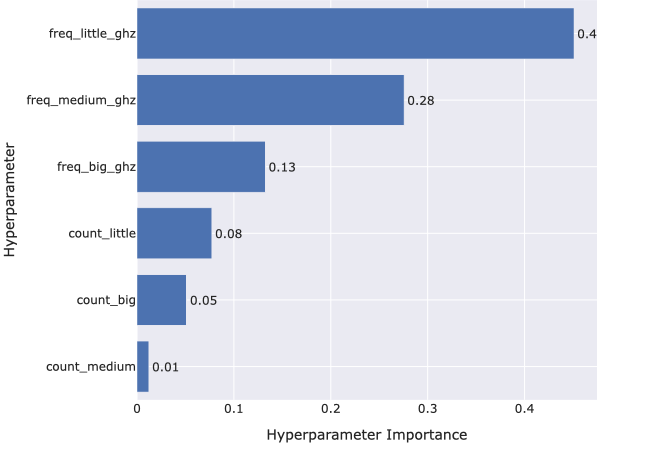}
        \caption{Sensitivity Analysis for RBF. }
        \label{fig:analy_kernel_rbf}
    \end{subfigure}
    
    \caption{Results of Sensitivity Analysis for different types of kernels. }
    \label{fig:analy_kernel}
\end{figure}

 Specifically, the Sensitivity Analysis in Figure~\ref{fig:analy_kernel} exposes a critical flaw in the RBF kernel. While the proposed Matérn 5/2 kernel (Figure~\ref{fig:analy_kernel_matern2.5}) correctly identifies \texttt{freq\_big\_ghz} and \texttt{freq\_medium\_ghz} as the dominant factors driving system performance, the RBF kernel (Figure~\ref{fig:analy_kernel_rbf}) yields a contradicting feature importance ranking, misleadingly attributing the highest sensitivity to \texttt{freq\_little\_ghz}. 

\par This discrepancy suggests that the RBF kernel's assumption of infinite smoothness causes it to misinterpret the causal relationship between hardware configurations and the objective function. The Matérn 5/2 kernel's ability to prioritize high-performance core frequencies aligns with the physical reality that latency is most sensitive to the clock speed of the most powerful computational units.

\subsection{Metric Evaluation}

\par
\begin{figure}[htbp]
    \centering
    \begin{subfigure}[b]{0.32\textwidth}
        \centering
        \includegraphics[width=\textwidth]{figures/analy_metric_balance.png}
        \caption{Sensitivity Analysis for Balance Metric Evaluation. }
        \label{fig:analy_metric_balance}
    \end{subfigure}
    \hfill
    \begin{subfigure}[b]{0.32\textwidth}
        \centering
        \includegraphics[width=\textwidth]{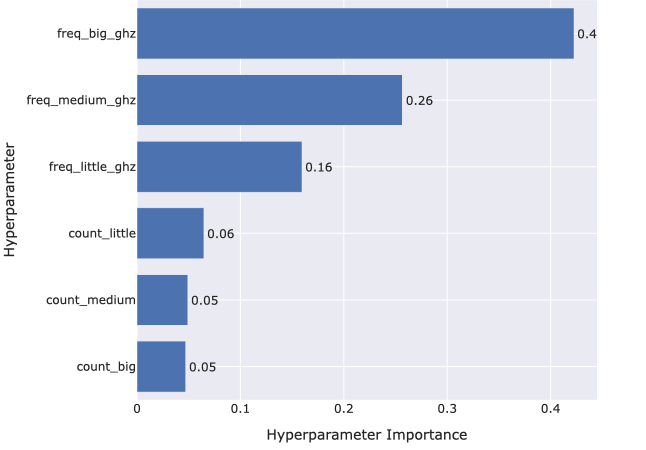}
        \caption{Sensitivity Analysis for Energy-Focused Metric Evaluation. }
        \label{fig:analy_metric_energy}
    \end{subfigure}
    \hfill 
    \begin{subfigure}[b]{0.32\textwidth}
        \centering
        \includegraphics[width=\textwidth]{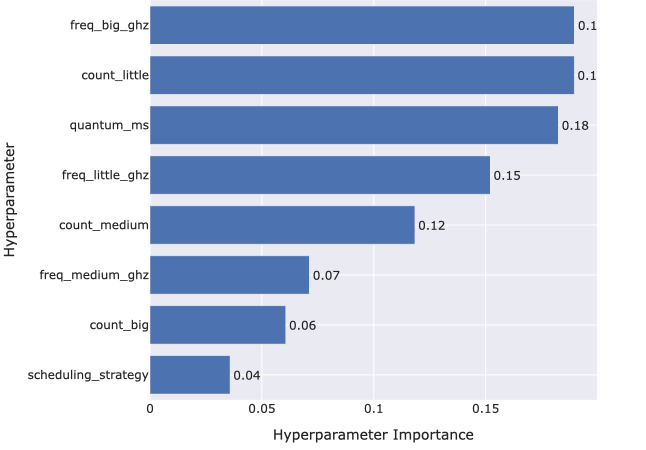}
        \caption{Sensitivity Analysis for Time-Focused Metric Evaluation. }
        \label{fig:analy_metric_time}
    \end{subfigure}
    
    \caption{Results of Sensitivity Analysis for different types of metrics. }
    \label{fig:analy_metric}
\end{figure}

 To assess the adaptability of the framework, we tested the optimizer under three distinct preference scenarios: \textit{Balance} ($\beta=\gamma=1$), \textit{Energy-First} ($\beta=3, \gamma=1$), and \textit{Time-First} ($\beta=1, \gamma=3$). Figure~\ref{fig:analy_metric} reveals the shift in hyperparameter importance.

\paragraph{Shift in Optimization Strategy}
The empirical results derived from the sensitivity analysis demonstrate that the optimizer effectively aligns the hardware configuration with high-level user preferences. By observing the variations in hyperparameter importance across different metrics (Figure~\ref{fig:analy_metric}), we identify distinct strategic adaptations:

\begin{itemize}[left=0.2cm]
    \item Time-Focused Scenario: 
    In the scenario where latency minimization is prioritized (Figure~\ref{fig:analy_metric_time}), the sensitivity analysis reveals that system performance is driven primarily by peak processing speed rather than parallel capacity. Consequently, \texttt{freq\_big\_ghz} exhibits the highest importance score. Interestingly, the model identifies \texttt{quantum\_ms} (scheduling time slice) and \texttt{count\_little} as critical factors, ranking them significantly higher than \texttt{count\_big}. This indicates that the optimizer prioritizes maximizing the clock speed of the primary core while fine-tuning the scheduler's granularity to minimize preemption overhead, rather than simply activating more big cores.

    \item Energy-Focused Scenario: 
    Conversely, under the energy-priority setting (Figure~\ref{fig:analy_metric_energy}), the relative importance of the big core count decreases as the model seeks to minimize static and dynamic power consumption. Instead, the optimizer directs its search effort toward tuning the medium core frequency and the little core count. This shift implies that the algorithm converges toward an efficiency-oriented strategy, where it attempts to offload tasks to the most energy-efficient cores or employs a "Race-to-Idle" strategy depending on the workload intensity. 
    
\end{itemize}

\begin{figure}[htbp]
    \centering
    \begin{subfigure}[b]{0.48\textwidth}
        \centering
        \includegraphics[width=\textwidth]{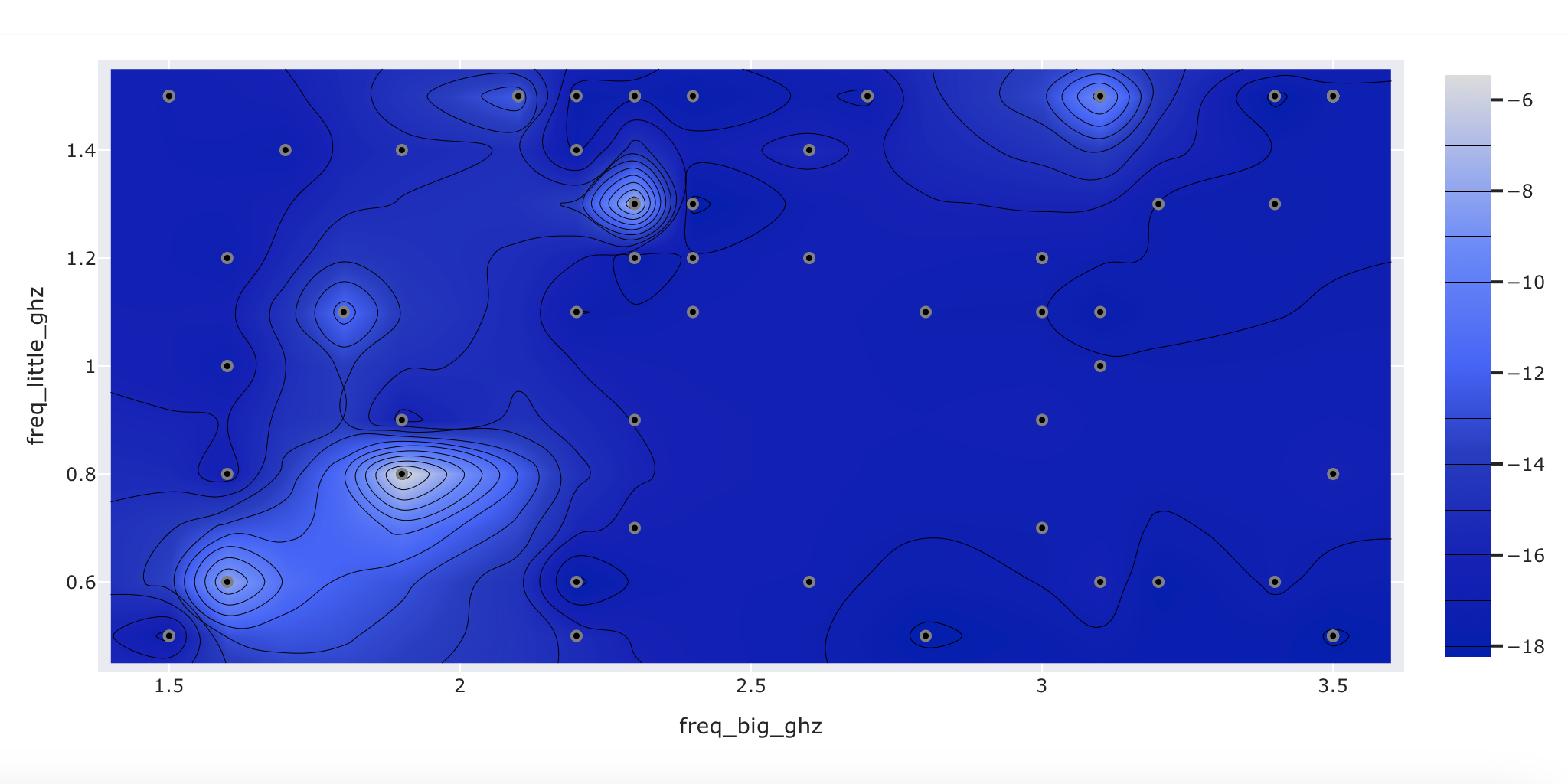} 
        \caption{Contour Plot of the Objective Value surface (Energy-Focused Scenario) visualizing the interaction between Big Core Frequency (X-axis) and Little Core Frequency (Y-axis). Darker regions indicate better performance (lower loss).}
        \label{fig:contour_energy}
    \end{subfigure}
    \hfill
    \begin{subfigure}[b]{0.48\textwidth}
        \centering
        \includegraphics[width=\textwidth]{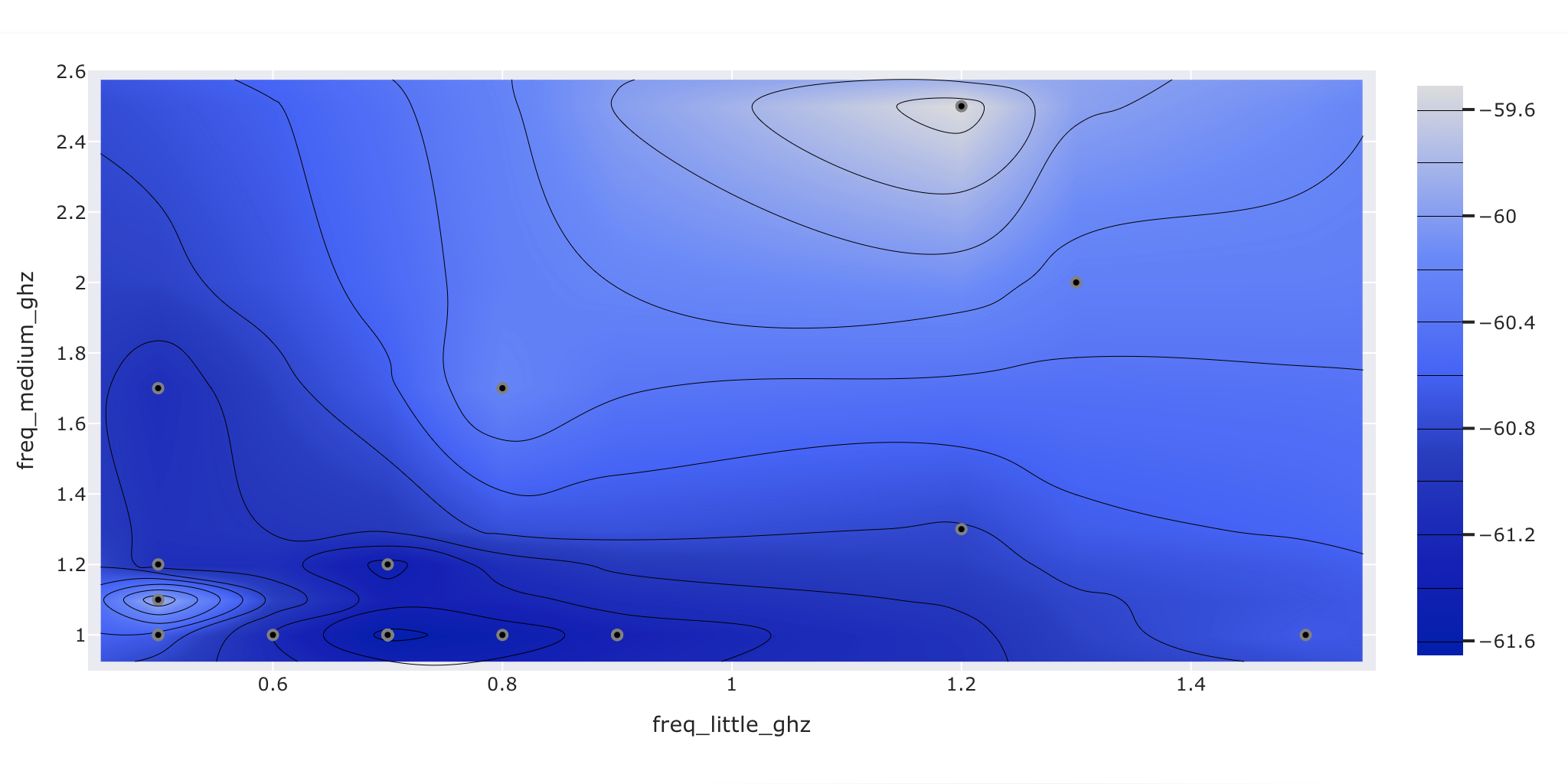}
        \caption{Contour Plot for Time-Focused Scenario (Little vs. Medium Frequency). Note the significantly narrower scale of the objective value compared to the energy scenario, indicating low sensitivity.}
        \label{fig:contour_time}
    \end{subfigure}
    \caption{Contour Plot for different types of metrics. }
    \label{fig:countour_metric}
\end{figure}

\paragraph{Physical Interpretation}
This dynamic shift in hyperparameter sensitivity confirms that the Bayesian Optimization framework captures the underlying physical trade-offs of the heterogeneous processor rather than simply memorizing data points. The algorithm automatically reallocates its exploration focus to different hardware resources in response to the specific definition of the loss function. This behaviour proves that the Gaussian Process surrogate model has successfully learned the distinct contribution of each core type toward energy efficiency and latency reduction, allowing it to navigate the design space intelligently without explicit human guidance.

\subsection{Robustness Test}

\par
\begin{figure}[htbp]
    \centering
    \begin{subfigure}[b]{0.32\textwidth}
        \centering
        \includegraphics[width=\textwidth]{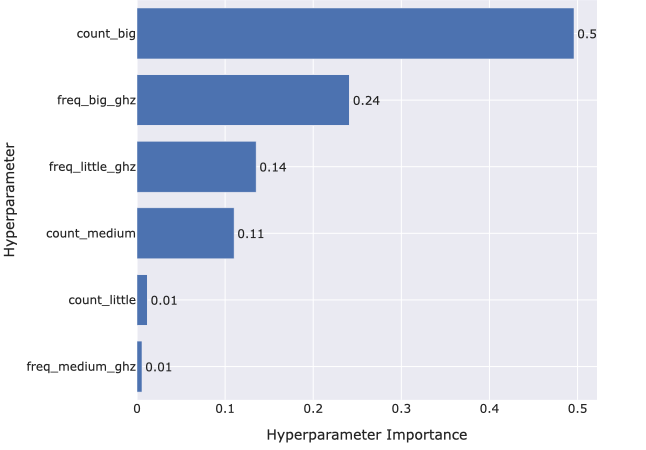}
        \caption{Sensitivity Analysis for $\lambda = 0.5$. }
        \label{fig:analy_lambda0.5}
    \end{subfigure}
    \hfill
    \begin{subfigure}[b]{0.32\textwidth}
        \centering
        \includegraphics[width=\textwidth]{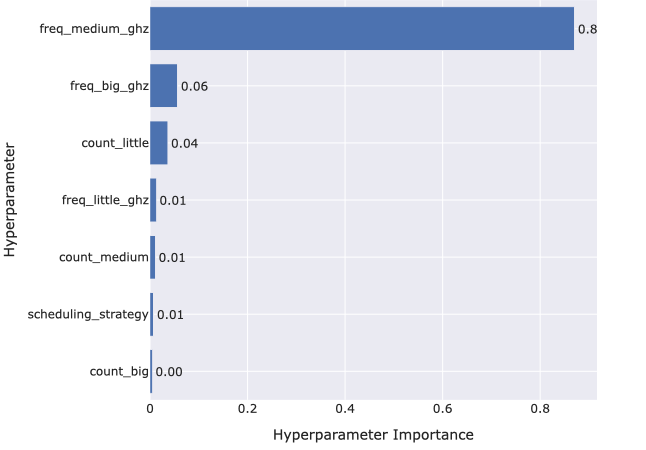}
        \caption{Sensitivity Analysis for $\lambda = 2.5$. }
        \label{fig:analy_lambda2.5}
    \end{subfigure}
    \hfill 
    \begin{subfigure}[b]{0.32\textwidth}
        \centering
        \includegraphics[width=\textwidth]{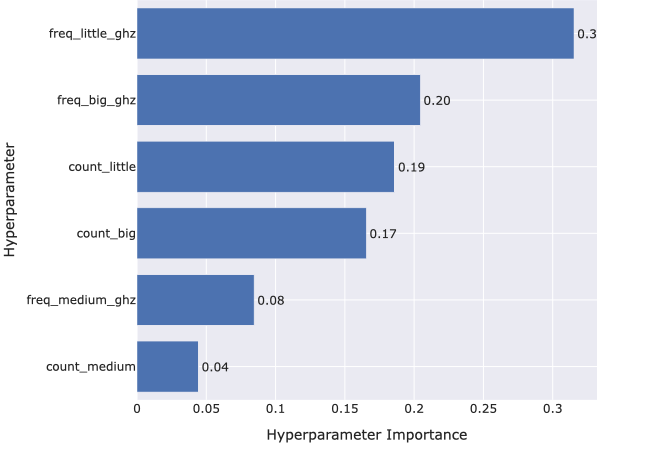}
        \caption{Sensitivity Analysis for $\lambda = 5.0$. }
        \label{fig:analy_lambda5.0}
    \end{subfigure}
    
    \caption{Results of Sensitivity Analysis for different lambda rates. }
    \label{fig:analy_lambda}
\end{figure}

 We evaluated the method robustness by testing under different workload intensities ($\lambda$) as visualized in Figure~\ref{fig:analy_lambda}.

\paragraph{Dynamic Scaling under Load}
The results confirm that the optimizer performs autonomous scaling strategies aligned with the workload pressure:

\begin{itemize}[left=0.2cm]
    \item Idle State ($\lambda=0.5$): 
    Under light load conditions, the Sensitivity Analysis (Figure~\ref{fig:analy_lambda0.5}) identifies \texttt{count\_big} as the overwhelmingly dominant factor (Importance $\approx 0.5$). This high sensitivity confirms that the optimizer correctly learns a strict constraint: Big Cores must be deactivated to minimize static power. Consequently, the system converges to a minimal configuration dependent on Little Cores, maximizing energy efficiency while maintaining acceptable latency.
    
    \item Full Scale-Out ($\lambda=2.5$): 
    This represents the system's optimal operating point. Facing high workload pressure, the Bayesian Optimizer fully activates the heterogeneous resources ($\mathbf{4 \times \text{Little}, 4 \times \text{Medium}, 4 \times \text{Big}}$). As shown in Figure~\ref{fig:analy_lambda2.5}, the focus shifts to \texttt{freq\_medium\_ghz}, indicating that once all cores are active, the optimizer fine-tunes the frequency of the most power-efficient clusters to balance the latency-energy trade-off.
    
    \item Throughput Saturation ($\lambda=5.0$): 
    Under extreme overload, the optimizer reverts to a minimal configuration dominated by Little Cores (Figure~\ref{fig:analy_lambda5.0}). While initially counter-intuitive, this behaviour is a mathematically optimal response to the logarithmic loss function. As queuing delays grow exponentially, the gradient of the latency term vanishes ($\lim_{T \to \infty} \frac{\partial \ln T}{\partial T} \approx 0$). Recognizing that performance targets are physically unattainable, the optimizer adopts a "damage control" strategy: it minimizes the energy penalty ($\ln E$) by throttling resources. 
\end{itemize}

\subsection{Multi-Objective Optimization}

\par In this final experiment, we treat Energy and Time as conflicting objectives to approximate the Pareto Frontier. To provide a clear visualization of the trade-off surface, outliers with extreme penalties were filtered from the scatter plot.

\begin{figure}[htbp]
    \centering
    % Left: Cleaned Pareto Front
    \begin{subfigure}[b]{0.48\textwidth}
        \centering
        \includegraphics[width=\textwidth]{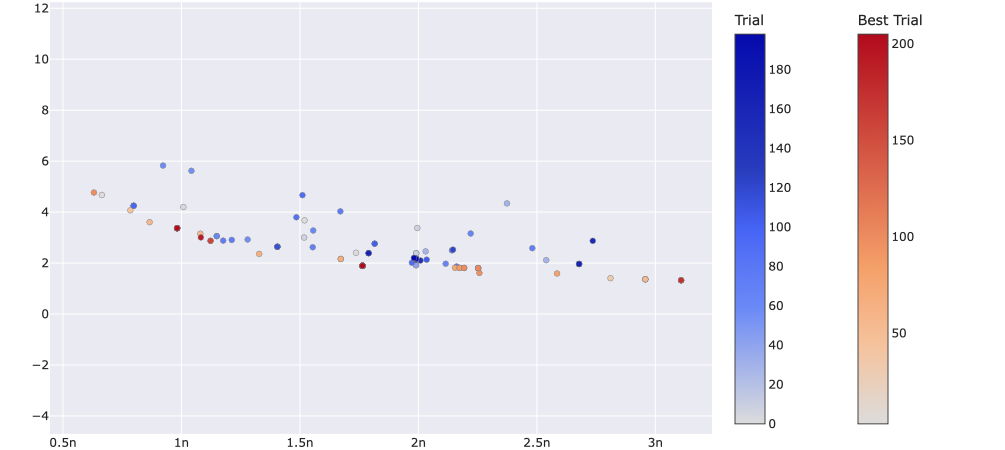} 
        \caption{Pareto Frontier Visualization.}
        \label{fig:moo_pareto}
    \end{subfigure}
    \hfill
    % Right: Importance Analysis
    \begin{subfigure}[b]{0.48\textwidth}
        \centering
        \includegraphics[width=\textwidth]{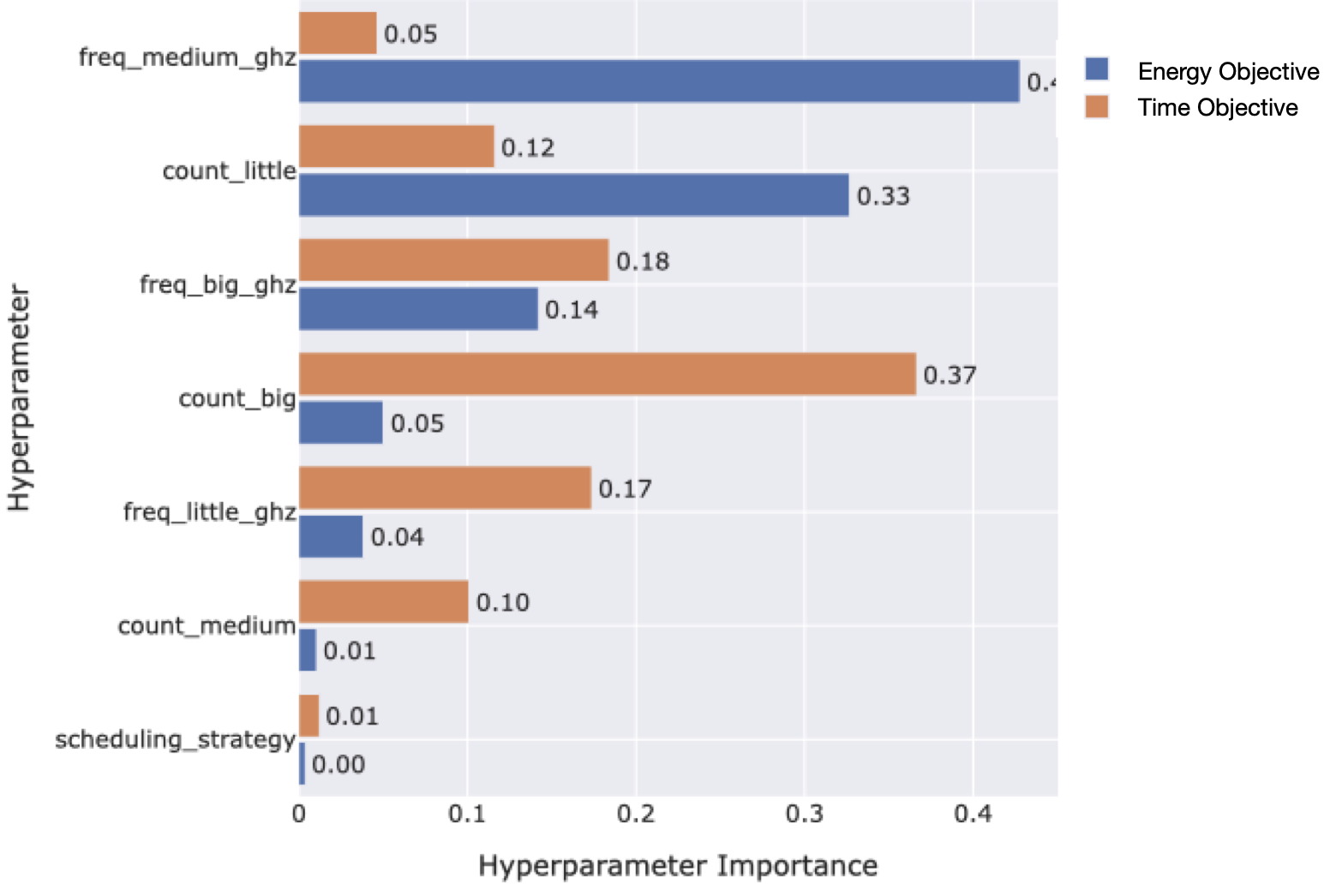} 
        \caption{Divergent Hyperparameter Importance.}
        \label{fig:moo_importance}
    \end{subfigure}
    
    \caption{Multi-Objective Optimization Results. (a) The scatter plot reveals the convex trade-off curve, where the "frontier" of non-dominated solutions represents the optimal compromise. (b) The importance analysis confirms the structural conflict: Objective 0 (Energy, blue) relies on Medium frequency and Little core count, while Objective 1 (Time, orange) is dominated by Big Core resources.}
    \label{fig:moo_results}
\end{figure}

\paragraph{Pareto Analysis}
The Multi-Objective Optimization results are summarized in Figure~\ref{fig:moo_results}.

\begin{itemize}[left=0.2cm]
    \item \textbf{Trade-off Discovery (Figure~\ref{fig:moo_pareto}):}
    The scatter plot clearly identifies the convex Pareto Frontier. We observe a dense cluster of non-dominated solutions ranging from low-energy/high-latency to high-energy/low-latency regions. The distinct curvature suggests the existence of a \textit{knee point}, representing a balanced configuration that avoids the performance extremes of both single-objective optimizations.

    \item \textbf{Structural Conflict (Figure~\ref{fig:moo_importance}):}
    The Hyperparameter Importance plot provides the physical explanation for this trade-off. We observe a clear divergence in resource sensitivity: The Energy objective (Blue) is primarily driven by \texttt{freq\_medium\_ghz} and \texttt{count\_little}, confirming that efficiency is maximized by tuning the mid-tier and low-power cores. In contrast, the Time objective (Orange) is overwhelmingly dominated by Big Core resources, specifically \texttt{count\_big} (Importance $\approx 0.37$) and \texttt{freq\_big\_ghz}. This stark separation proves that the MOO algorithm successfully decoupled the conflicting physical requirements of the heterogeneous system.
\end{itemize}

    \section{Conclusion} \label{sec:conclusion}

\par In this work, we reformulated the scheduling parameter tuning of heterogeneous multi-core systems as a Bayesian Optimization problem. Beyond demonstrating the algorithmic efficiency of the framework, our extensive sensitivity and multi-objective analyses revealed three critical insights into the physical behaviour of post-Dennard architectures:

\begin{enumerate}[left=0.2cm]
    \item \textbf{Non-Smoothness of the Performance Landscape:} The superior performance of the Matérn 5/2 kernel over the RBF kernel confirms that the optimization landscape is characterized by sharp performance cliffs rather than smooth gradients. This necessitates the use of roughness-aware surrogate models to handle discrete resource discontinuities.
    
    \item \textbf{Emergence of "Race-to-Idle" and Resource Decoupling:} The optimizer autonomously discovered sophisticated control strategies without expert heuristics. Specifically, it validated the "Race-to-Idle" theory in energy-constrained scenarios (activating high-frequency big cores to minimize leakage) and demonstrated a structural decoupling where latency optimization is governed by big cores while energy efficiency is managed by little cores.
    
    \item \textbf{Phase Transition under Saturation:} Our robustness tests identified a distinct phase transition in the optimal strategy. While resource scaling is effective in the linear workload region ($\lambda \le 2.5$), the system shifts to a "penalty mitigation" mode under saturation ($\lambda=5.0$), minimizing energy consumption when throughput limits render latency optimization impossible.
\end{enumerate}

\paragraph{Future Work} Although this study successfully demonstrates the efficacy of Bayesian Optimization for offline parameter tuning, several avenues remain for extending the framework's applicability to dynamic, real-world environments. 

First, future research will focus on transitioning from static offline configuration to online run-time adaptation. By integrating lightweight surrogate models, such as Contextual Bandits, the system could dynamically adjust voltage and frequency (DVFS) settings in response to real-time traffic bursts, rather than relying on a fixed schedule. 

Additionally, we aim to relax the assumption of independent tasks by incorporating task dependency models, specifically Directed Acyclic Graphs (DAGs). Handling inter-task dependencies introduces additional challenges regarding communication overhead and pipeline stalling, which are critical for complex embedded applications like video processing.

% \todo[inline]{to be updated.}

    \newpage
    
    \bibliographystyle{unsrt}  
    \bibliography{references}  

    \newpage

    % APPENDIX NOT NEEDED FOR FINAL SUBMISSION
    \appendix
    % \newpage
    \section{Details and additional results}

% \section{Images}

\begin{figure}[htbp]
    \centering
    \begin{subfigure}[b]{0.32\textwidth}
        \centering
        \includegraphics[width=\textwidth]{figures/res_metric_balance.png}
        \caption{Optimization History for Balance Metric Evaluation. }
        \label{fig:res_metric_balance}
    \end{subfigure}
    \hfill
    \begin{subfigure}[b]{0.32\textwidth}
        \centering
        \includegraphics[width=\textwidth]{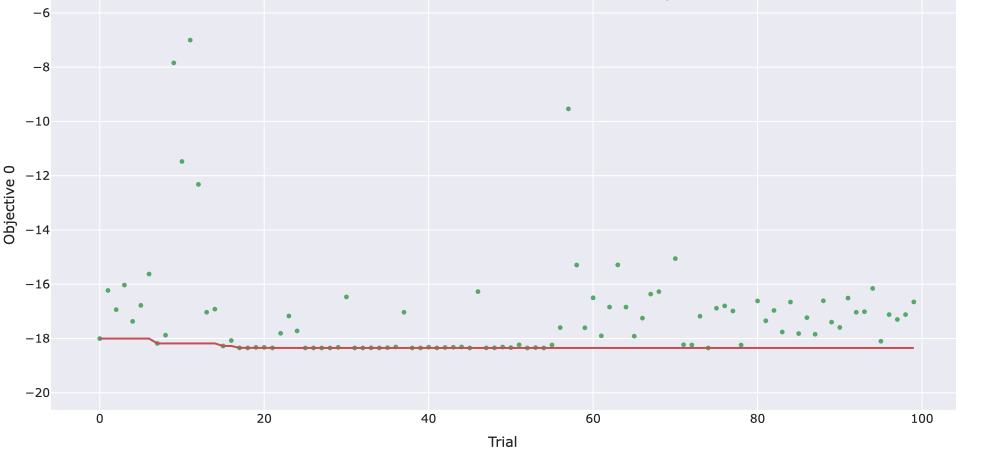}
        \caption{Optimization History for Energy-Focused Metric Evaluation. }
        \label{fig:res_metric_energy}
    \end{subfigure}
    \hfill 
    \begin{subfigure}[b]{0.32\textwidth}
        \centering
        \includegraphics[width=\textwidth]{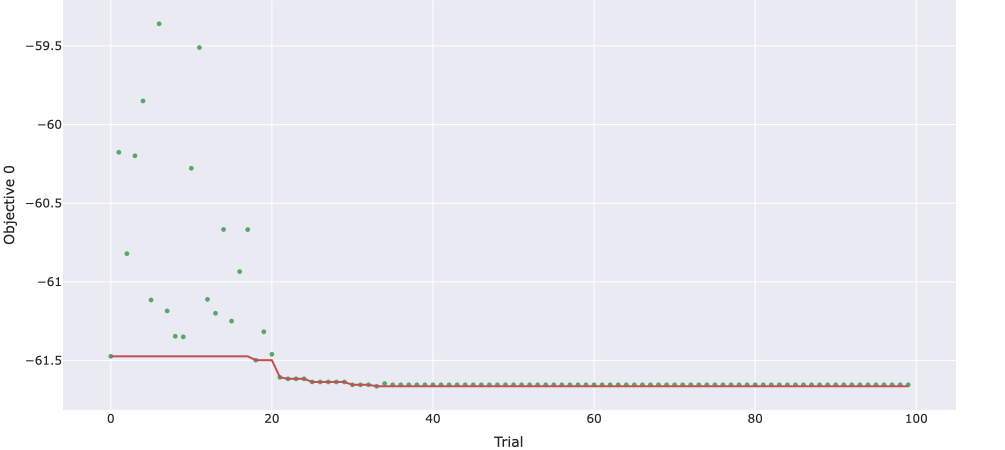}
        \caption{Optimization History for Time-Focused Metric Evaluation. }
        \label{fig:res_metric_time}
    \end{subfigure}
    
    \caption{Results of Bayesian Optimization history for different types of metrics. }
    \label{fig:res_history_metric}
\end{figure}

\begin{figure}[htbp]
    \centering
    \begin{subfigure}[b]{0.32\textwidth}
        \centering
        \includegraphics[width=\textwidth]{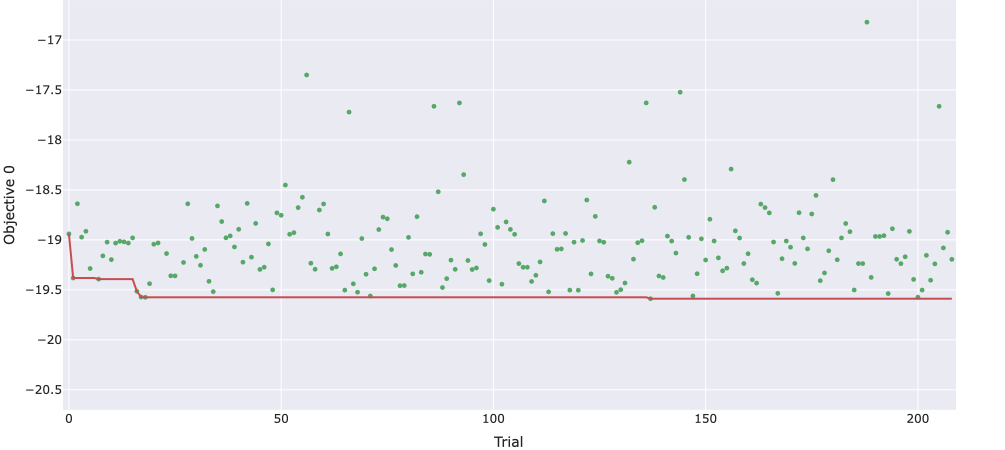}
        \caption{Optimization History for $\lambda = 0.5$. }
        \label{fig:res_lambda0.5}
    \end{subfigure}
    \hfill
    \begin{subfigure}[b]{0.32\textwidth}
        \centering
        \includegraphics[width=\textwidth]{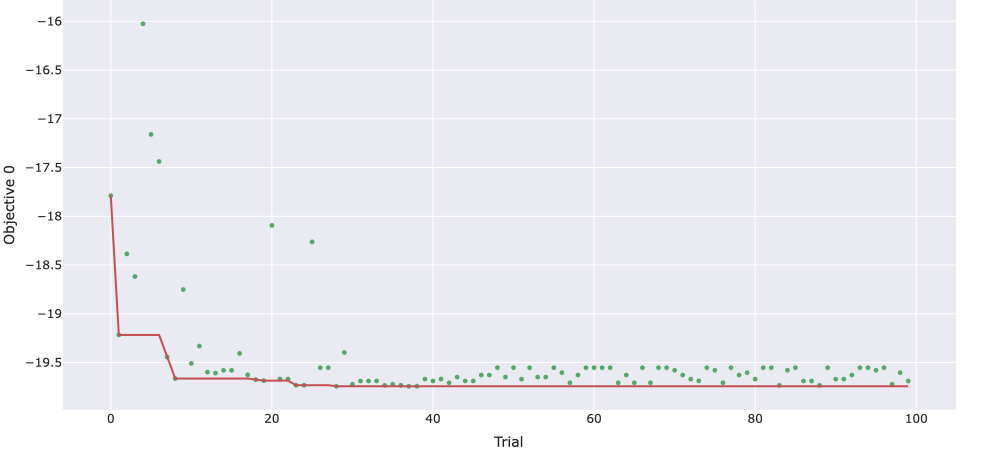}
        \caption{Optimization History for $\lambda = 2.5$. }
        \label{fig:res_lambda2.5}
    \end{subfigure}
    \hfill 
    \begin{subfigure}[b]{0.32\textwidth}
        \centering
        \includegraphics[width=\textwidth]{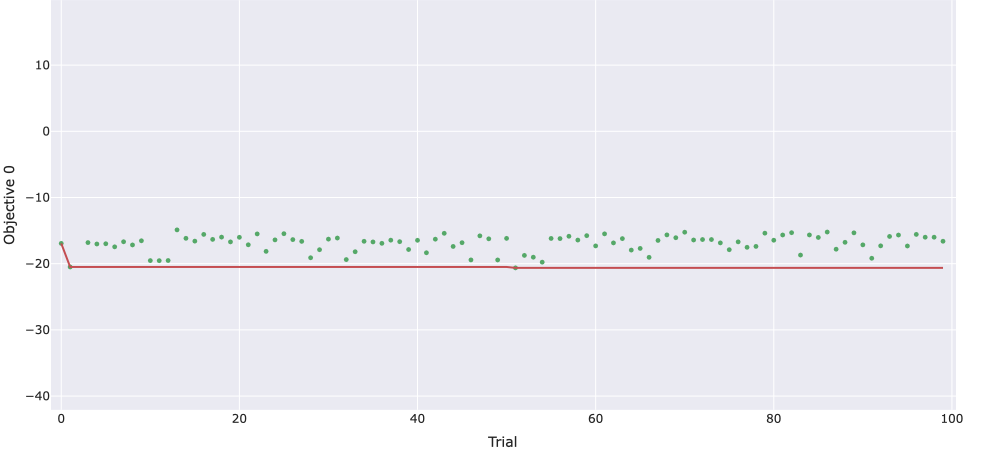}
        \caption{Optimization History for $\lambda = 5.0$. }
        \label{fig:res_lambda5.0}
    \end{subfigure}
    
    \caption{Results of Bayesian Optimization history for different lambda rates. }
    \label{fig:res_lambda}
\end{figure}  % optional
    \section{Proposal} \label{sec:proposal}

\paragraph{Main goals.} We use ML to learn knowledge regarding how CPU scheduling affects the efficiency–energy trade-off. Specifically, it covers the following aspects, 

- Which aspects of scheduling behaviour (e.g., pre-emption frequency, load balance, task affinity) most influence efficiency and energy use.

- How tuneable conventional scheduling or hybrid algorithms (e.g., priority-based, load-balancing, or DVFS-aware) can optimise throughput, latency, and power consumption; 

- How ML can learn knowledge and thus provide insights into scheduling decisions, e.g. enhancing traditional heuristics by learning parameters, predicting workloads, or guiding real-time decisions.
           % optional
    % NOT NEEDED FOR FINAL SUBMISSION
% \section{Individual contribution report}

% \paragraph{Instructions.}

% \textit{While this is a group project we are also required by the department to collect information on what each individual have done for the project. To do so each group member should submit a up to one page document with the following content.}

% - the CSRid of all members in your group.

% - summary of your personal contribution to the work of the team.

% - your assessment of the contributions made by each other member of the team.

% \paragraph{Members.} Peter Hu (zh369), Yifei Shi (ys690).

% 
\section{Author contributions} Both authors jointly discussed and researched relevant prior work, assisted each other with debugging, and participated in reviewing and improving the overall codebase.

\paragraph{Personal contribution.} 

\paragraph{Zheyuan (zh369)}

\begin{itemize}[left=0.2cm]
\item Implemented \texttt{SimPy} simulation frameworks for scheduling algorithms, including,
    \begin{itemize}[left=0.2cm]
      \item First-Come, First-Served (FCFS),
      \item Time-sliced / quantum based,
      \item Round Robin (RR),
      \item Priority-based scheduling.
    \end{itemize}
\item Designed and implemented the task models used in the project.
\item Implemented detailed logging of scheduler behavior, tracking task attributes, task assignments, core energy consumption, latency, fairness and throughput.
\item Researched the related metrics with underlying theory, e.g. active or idle power consumption w.r.t. frequency. Implemented per-run evaluation of energy consumption and performance metrics.
\item Contributed to overall system design discussions, including the optimization component among,
    \begin{itemize}[left=0.2cm]
        \item Gaussian Processes (GP),
        \item different kernel choices,
        \item Bayesian Optimization (BO),
        \item Sensitivity Analysis.
    \end{itemize}
\item Authored the Abstract, Related work, Simulation sections of the paper, including the underlying motivations, theories and practical design choices. 
\item Refined the Introduction section for our experimental settings.
\item Contributed to overall paper editing, including figures, tables, and result interpretation.
\end{itemize}

\paragraph{Yifei (ys690)}
\begin{itemize}[left=0.2cm]
\item  Implemented the optimization component from simulator outputs of latency, throughput, fairness, and energy.
\item Researched, designed overall methodologies, including,
    \begin{itemize}[left=0.2cm]
        \item Gaussian Processes (GP),
        \item different kernel choices,
        \item Bayesian Optimization (BO),
        \item Sensitivity Analysis.
    \end{itemize}
\item Contributed to the scheduling simulators, including,
    \begin{itemize}[left=0.2cm]
      \item First-In, First-Out (FIFO),
      \item Time quantum (or slice) based,
      \item Round Robin (RR),
      \item Priority-based scheduling.
    \end{itemize}
\item Designed and contributed to the task models used in the project, e.g. modelling tasks as a Poisson process.
\item Design modular experiment codebase to support flexible settings for Bayesian Optimization controlled by \texttt{.yaml} files.
\item Override native interface of \texttt{optuna} library to support the replacement of kernels. 
\item Authored the Introduction, Emulation, Discussion sections of the paper.
\item Refined the other sections.
\item Contributed to overall paper editing, including figures, tables, and result interpretation.
\end{itemize}

% \paragraph{Assessment of the contributions.}

\paragraph{Acknowledgement.} \label{sec:ack}

We sincerely appreciate Professor Carl Henrik Ek for organizing the exciting module L48 Machine Learning and the Physical World at Department of Computer Science and Technology, University of Cambridge and providing consistent feedback regarding this project during the proposal and viva phase.

\end{document}